\documentclass[twocolumn,amsmath,amssymb,aps,pre]{revtex4}
\usepackage{xcolor}
\usepackage{graphicx}
\usepackage{colordvi,epsfig,bm,hyperref}
\usepackage[utf8]{inputenc}
\usepackage{mathtools}
\usepackage{comment}  
\usepackage{natbib}
\usepackage{siunitx}
\usepackage{graphicx}
\usepackage{float}

\begin{document}

\title{Overcoming the complexity barrier of the dynamic message-passing method in networks with fat-tailed degree distributions}

\author{Giuseppe Torrisi }
 \email{giuseppe.torrisi@kcl.ac.uk}
\author{Alessia Annibale}%
  \author{Reimer K\"uhn}
\affiliation{%
Department of Mathematics, King's College London, Strand,\\ London WC2R 2LS, United Kingdom 
}%

\date{\today}

\begin{abstract}
The dynamic cavity method provides the most efficient way to evaluate probabilities of dynamic trajectories in systems of stochastic units with unidirectional sparse interactions. It is closely related to sum-product algorithms widely used to compute marginal functions from complicated global functions of many variables, with applications in disordered systems, combinatorial optimization and computer science. However, the complexity of the cavity approach grows exponentially with the in-degrees of the interacting units, which creates a de-facto barrier for the successful analysis of systems with fat-tailed in-degree distributions. In this manuscript, we present a dynamic programming algorithm that overcomes this barrier by reducing the 
computational complexity in the in-degrees 
from exponential to quadratic, whenever couplings are chosen randomly from (or can be approximated in terms of) discrete, possibly unit-dependent, sets of equidistant values. As a case study, we analyse the dynamics of a random Boolean network with a fat-tailed degree distribution and fully asymmetric binary $\pm J$ couplings, and we use the power of the algorithm to unlock the noise dependent heterogeneity of stationary node activation patterns in such a system. 
\end{abstract}

\maketitle
\section{Introduction}
The collective behaviour of a broad range of disordered and complex systems can be analysed in terms of models of interacting binary units, evolving according to stochastic Boolean functions of the neighbouring units. They include spin glasses \cite{EdwardsAnderson75, SK75, viana1985phase} in physics, gene-regulatory and immune networks \cite{kauffman1969metabolic, derrida1986random, ParisiPNAS1990, AgliariEtAl2012, agliari2013immune} in biology, artificial neural networks \cite{Hopfield82, hertz2018introduction, Amit87a, sollich2014extensive}
in computer science, agent based models \cite{ChalletZhang97, CoolenMG05, Iori99, Bornh01}, and models of operational or credit risk \cite{ AnandKuehn07, HaKu06} in economics and finance, and a variety of hard combinatorial optimization problems \cite{weigt2000number, cocco2001statistical, martin2001statistical, Franz+01, mezard2002analytic} to name but a few.

Whenever the system dynamics satisfies detailed balance, equilibrium statistical mechanics provides a powerful arsenal of techniques to analyse their operation at stationarity.  However, 
when the dynamics does not satisfy detailed balance -- as in systems that are driven, dissipative or exhibit a degree of asymmetry in the interactions -- or 
when one is interested in genuine dynamical aspects of the problem, one has to resort to tools of non-equilibrium statistical mechanics. These are typically much more cumbersome to use, involving e.g. non-Markovian dynamics of an ``effective'' local degree of freedom which requires solving dynamic self-consistency equations for global correlation and response functions in the case of fully connected systems \cite{Dominicis78, SommersPRL1987}. In systems with parallel dynamics, the complexity of the analysis grows exponentially in the number of time steps considered \cite{gardner1987zero} and quickly becomes infeasible. If systems exhibit fully asymmetric interactions though, the analysis simplifies considerably \cite{crisanti1988dynamics} 
and can often be analysed in some explicit detail. 

Techniques for fully connected systems carry over to 
systems with diluted interactions if the mean connectivity diverges in the large system limit \cite{KreeZipp91,Derrida87}. More recently, however, there has been considerable interest in the dynamics of genuinely sparse systems, defined on complex networks which remain {\em finitely coordinated\/} in the thermodynamic limit \cite{DorogMend03, NewBk10, Estrada2011, Latora+2017}. In these systems, approximations schemes  successfully used for dense systems, such as heterogeneous  mean-field and TAP approaches \cite{mezard2002analytic,roudi2011dynamical}, have been shown to be ineffective \cite{zhang2012inference,aurell2012dynamic}. On the other hand, 
the dynamic cavity method  \cite{mimura2009parallel,neri2009cavity, PagKu15, lokhov2014inferring} is suitable to analyse sparse systems, where short loops are rare. This method 
requires to follow the evolution of a collection of {\em local\/} dynamical variables, whose number is proportional to system size, in a {\em cavity} graph, where, however, 
the dynamics is non-Markovian 
whenever there is some degree of symmetry in the interactions. This leads to the aforementioned exponential scaling with the time horizon considered. 
Only when interactions are fully asymmetric, 
those retarded self-interactions that make dynamics non-Markovian \cite{Derrida87, derrida1987dynamical, mimura2009parallel, neri2009cavity}
are absent and the exponential time 
complexity is removed.
For this reason, systems with fully asymmetric interactions, that can be analysed within a Markovian framework, are a particularly attractive option to use for a first level of analysis.
However,
for a large class of models (including linear and non-linear threshold models) widely used in physics, biology, finance and social science, the evaluation of time dependent averages of local dynamical variables
grows {\em exponentially\/} with the in-degree of the node \cite{mimura2009parallel, neri2009cavity}, even in the Markovian case. For a wide variety of systems of topical interest, which are characterized by fat-tailed degree distributions, (see e.g. \cite{Lesk14, AlbBarab02, DorogMend03, NewBk10, Estrada2011, Latora+2017} for general background), this
creates an additional and seemingly impenetrable complexity barrier that effectively prevents any analytic study of such systems.
This barrier is also present in the context of spreading processes, see ref. \cite{altarelli2013optimizing}, where the authors  
describe an optimisation tool that reduces the exponential complexity to polynomial. However, their result relies on the microscopic irreversibility of the dynamics in the sense that a node that becomes active at a certain time will never revert to inactive state. In contrast, the dynamics investigated in the present manuscript is not irreversible, due to the presence of both positive and negative interaction terms. Hence, the dynamics cannot be solved via the methods used in Refs\,. \cite{altarelli2013optimizing, lokhov2015dynamic}, and a new method is therefore required.

The main aim of the present contribution is to provide a way to overcome this complexity barrier. We propose an algorithm inspired by dynamic programming that reduces the dependence of the computational complexity on the in-degrees of the system from exponential to quadratic. The approach can be shown to work for spin systems and Boolean networks, including those with $p$-node interactions, and is applicable whenever couplings are chosen randomly from a discrete set of equidistant values that may well be site-dependent. This would carry over to systems exhibiting continuous couplings that can be well approximated in this way, e.g. by discretization. To the best of our knowledge, our work represent the first to crack the complexity barrier in the degree for microscopically reversible dynamics. For the sake of definiteness, we shall explain it for the case of networks of Boolean linear threshold units, and will cover generalisations to other systems in an extended paper. The interested reader can find the code to reproduce the results shown in the present paper at the following link \footnote{ \url{https://zenodo.org/badge/latestdoi/355104547}}.

Our paper is organized as follows. In Sec.\,II we describe the model that is being studied. Sec.\,III is devoted to the dynamic programming approach devised to overcome the exponential complexity barrier. In Sec.\, IV  we demonstrate our method on an application using a network with a fat-tailed degree distribution. Sec.\,V finally contains a conclusion and discussion. We also included an appendix which contains details concerning numerics and simulations. 

\section{The Model}
\label{sec:model}
We consider a Boolean network model defined on a finitely connected directed graph of $N$ nodes labelled $i=1,\dots, N$. For each edge $(ij)$ of the graph, the edge weight $J_{ij}\in \mathbb{R}$ defines the strength of the interaction carried from node $j$ to node $i$. Here we assume that the non-zero edge weights are binary, and randomly and independently drawn from the distribution $P(J_{ij})=\eta \delta(J_{ij}-J) + (1-\eta)\delta(J_{ij}+J)$ with $\eta \in (0,1)$. We  use $\partial_i^{in}=\lbrace j: J_{ij}\neq 0\rbrace$ to denote the set of predecessors of node $i$ and use $k_i= |\partial_i^{in}|$ to designate the in-degree of node $i$. With each node we associate a Boolean state variable $n_i=\lbrace 0,1\rbrace$.  Each node $i$ receives a signal given by the local field 
\begin{equation}
    h_i\big(\bm{n}(t)\big) = \sum_j J_{ij}\, n_j(t)
    \label{eq:field}
\end{equation} 
and determines its activation state at the next time step following a noisy linear threshold dynamics of the form
\begin{equation}
    n_i(t+1)=\Theta \big[h_i\big(\bm {n}(t)\big) -\vartheta_i -z_i(t)\big]\ .
    \label{eq: model}
\end{equation}
Here the $\vartheta_i$ are local thresholds, and the $z_i(t)$ are independent identically distributed random variables extracted,
from some cumulative distribution function $\mbox{Prob}[z \le x]=\Phi(x)$. A common choice for the noise is `thermal' noise described by the distribution
$\Phi(x) =\Phi_\beta(x) = [1+\tanh(\beta x/2)]/2$ in which $\beta^{-1} =T$ is a measure of the noise strength. Any other noise model could be chosen instead. The local field $h_i(\bm{n})$ depends only on the states of those nodes which are predecessors of node $i$, that we denote with $\bm{n}_{\partial_{i}}=\lbrace n_j ,j \in{\partial_i^{in}}\rbrace$. In the following, we will thus 
write the local field in $i$ as  $h_i(\bm{n}_{\partial_{i}})$. 
 
The dynamics described by Eqs.\,\eqref{eq:field}  and \eqref{eq: model} is stochastic. Let us denote by $P_i(t) = \mbox{Prob}(n_i(t)=1)$ the activation probability of node $i$ at time $t$. For finitely coordinated random graphs of the type considered here, the cavity method \cite{mezard2001bethe} can be used to analyse the dynamics of the system. It is exact on trees and known to become exact for finitely coordinated random graphs in the thermodynamic limit, as the typical length of any loops in such systems diverges logarithmically in system size $N$. Using the cavity approach on a fully-asymmetric network, see \cite{mimura2009parallel,neri2009cavity,aurell2012dynamic},  a closed form expression for the activation probability $P_i(t+1)$ is obtained as \begin{equation} P_i(t+1)=
\left\langle \Phi_\beta\left(h_i(\bm{n}_{\partial_{i}})-\vartheta_i \right)\right\rangle_{ \bm{n}_{\partial_{i}},\,t}\ ,
\label{eq: result cavity}
\end{equation}
in which the angled bracket $\langle\,\cdot\,\rangle_{ \bm{n}_{\partial_{i}},\,t}$ indicates an average over the states of the predecessors of node $i$, $\bm{n}_{\partial_{i}}$
evaluated using their joint node activation probabilities at time $t$.
\begin{equation}
\langle(\cdot)\rangle_{ \bm{n}_{\partial_{i}},\,t}=\sum_{ \bm{n}_{\partial_{i}}}(\cdot) \prod_{j\in\partial_i^{\mathrm{in}}}P_j(t)^{n_j}[1-P_j(t)]^{1-n_j}\,.
\label{eq:average_definition}
\end{equation}
Within the cavity approximation these joint probabilities can be taken to be independent. However, in order to evaluate the average, one has to sample from a state space containing $2^{k_i}$ elements for a node with in-degree $k_i$.  This creates the aforementioned exponential complexity barrier for systems exhibiting fat-tailed in-degree distributions.  In order to overcome it, we exploit the independence of joint node activation probabilities of a given node assumed within the cavity formalism to devise a recursive algorithm inspired by dynamic programming that is able to reduce the computational complexity of evaluating local averages from exponential to polynomial in the in-degree of nodes.  We note that Eq.\, \eqref{eq: result cavity} is of the form 
\begin{equation}
    P_i(t+1) = \sum_{\bm{n}_{\partial_{i}}}\phi_i\left(h_i(\bm{n})\right)\prod_{j\in \partial_i^{\mathrm{in}}}\psi_j(n_j)
\label{eq:generalised_cavity}
\end{equation}
with $\phi_i(x) = \Phi_\beta \big(x-\vartheta_i\big)$ and $\psi_j (x) = P_j(t)^x(1-P_j(t))^{1-x}$.
This structure of the cavity equation (Eq.\, \eqref{eq:generalised_cavity}) is shared by belief-propagation involving sum-product algorithms, as broadly used in information theory and machine learning
\cite{mezard2009information}. Our method tackles the exponential complexity barrier also in those contexts.

\section{Dynamic programming}  
\label{sec:dyn-pro}
To introduce the idea behind the recursive dynamic programming approach \cite{bellman1966dynamic, scott2009polynomial} in the present context, let us consider a node $i$ with in-degree $k_i$, and let us use $\lbrace 1,\dots k_i\rbrace$ to label the indices of the predecessors of node $i$.
One way is to evaluate $P_i(t+1)$ iteratively. That is, one first evaluates the average over $n_1$ and obtains 
 \begin{eqnarray}
 P_i(t+1)\! &=&\! P_1(t) \left\langle\! \Phi_\beta \Bigg(J_{i1}+\sum_{j=2}^{k_i}J_{ij} n_j\!-\!\vartheta_i\Bigg) \right\rangle_{n_{2,\dots,k_i},t}\nonumber\\
& &\!\!\!+(1-P_1(t)) \left\langle\! \Phi_\beta \Bigg(\sum_{j=2}^{k_i}J_{ij}n_j \!-\!\vartheta_i\Bigg)\right\rangle_{n_{2,\dots,k_i}, t}\ ,
 \label{eq:dynamic programming}
 \end{eqnarray}
in which $\langle\,\cdot\,\rangle_{n_{2,\dots,k_i},t}$ denotes averages over $n_2,\dots,n_{k_i}$ that remain to be performed. The original problem involving $k_i$ predecessors has thus been reduced to {\em two\/} smaller problems, each requiring to perform averages over $k_i-1$ variables.  This procedure is repeated recursively on each of the smaller problems until all averages are evaluated, which overall requires to evaluate $2^{k_i}$ averages for a vertex of degree $k_i$.

In order to benefit from the heuristics of dynamic programming, we propose to reformulate the evaluation of the average Eq.\,\eqref{eq: result cavity} through the recursion just described, by {\em first solving a more general problem}, namely the problem to evaluate the {\em family of averages\/}  $\{f_i(\ell,\tilde{h}); 1\leq \ell\leq k_i; \tilde h \in \mathbb{R}\}$ defined by
\begin{equation}
      f_i(\ell,\tilde{h}) = \left\langle\! \Phi_{\beta}\Bigg(\tilde{h}+\sum_{j=\ell}^{k_i}J_{ij}n_j -\vartheta_i\Bigg) \right\rangle_{n_{\ell,\dots,k_i}, t}\ .
\end{equation}
We will refer to $\tilde h$ as an auxiliary field.
At any given $\ell$, the set of $f_i(\ell,\tilde{h})$ of interest would in fact correspond to the set of averages representing the sub-problems that remain to be evaluated after $\ell-1$ levels of the above recursive procedure have been performed.
With reference to Eq.\,\eqref{eq:dynamic programming}, the $f_i(\ell,\tilde{h})$ are obtained using the {\em backward recursion} \begin{eqnarray}
   f_i(\ell,\tilde{h}) &=& P_\ell(t)\, f_i(\ell+1,\tilde{h} +J_{i\ell})\nonumber\\
   & & +\big(1-P_\ell(t)\big)\,f_i(\ell+1,\tilde{h}) 
   \label{eq: recursive}
\end{eqnarray}
for $1\leq \ell\leq k_i$, with the {\em terminal boundary condition}
\begin{equation}
  f_i(k_i+1,\tilde{h}) = \Phi_{\beta}\left(\tilde{h} -\vartheta_i\right)\ .
\end{equation}

The original average $P_i(t+1)$ of  Eq.\,\eqref{eq: result cavity} that we are ultimately interested in is, within this backward iteration scheme, obtained as
\begin{equation}
    P_i(t+1) = f_i(1,0)\ .
\end{equation}

\begin{figure}
 \includegraphics[width = 0.475\textwidth]{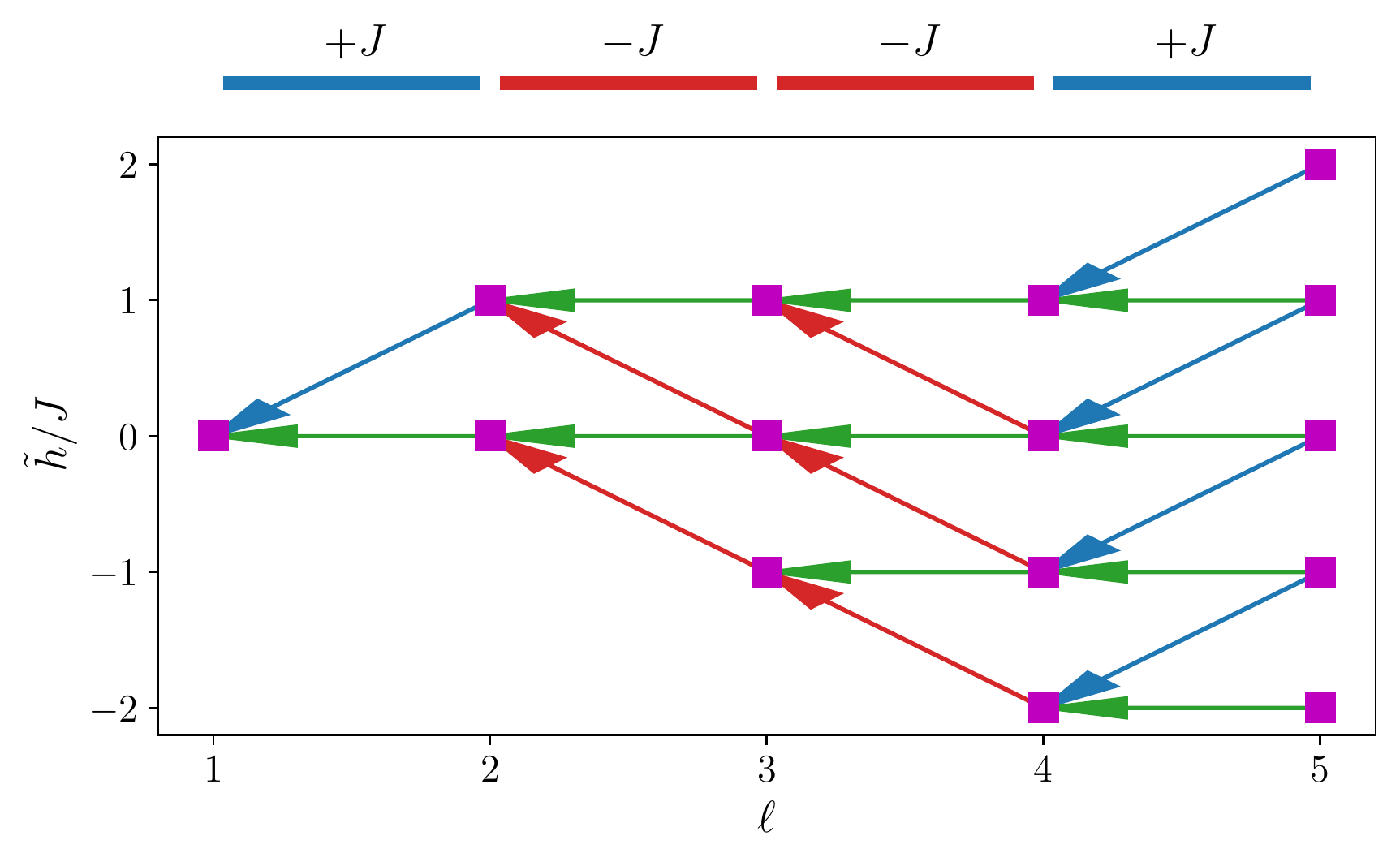}~~~~~
 \caption{Set of $(\ell,\tilde{h})$ combinations for which  $f(\ell,\tilde{h})$ is required in the execution of recursive evaluation Eq.\,\eqref{eq: recursive} for a particular choice of couplings. In this example, $k_i = 4$, the vector of couplings is $(J_{i,1},J_{i,2},J_{i,3},J_{i,4}) = (+J,-J,-J,+J)$ as shown on the top. Arrows joins   configurations  $(\ell,\tilde{h})$ explored through the  recursive relation Eq.\, \eqref{eq: recursive} . Green arrow indicates the recursion  call from  $f_i(\ell+1,\tilde{h})$ to  $f_i(\ell,\tilde{h})$. Blue  and red arrows indicates the call from  $f_i(\ell+1,\tilde{h}+J_{il})$ to $f_i(\ell,\tilde{h})$ through a positive or negative coupling respectively.}
 \label{fig:dynamic programming}
 \end{figure}

 While this recursive formulation is equivalent to direct evaluation of Eq.\,\eqref{eq: result cavity} and does not provide any speedup per se, the evaluation using dynamic programming implemented in the recursive procedure  Eq.\,\eqref{eq: recursive} leads to a dramatic reduction of computational complexity if the non-zero 
weights $J_{i\ell}$ of the incoming edges to site $i$
are chosen in a discrete set of equidistant values, which may even be site-dependent, i.e. $J_{ij}\in\{-r_i J_i,\ldots,-J_i,0,J_i,\ldots, s_iJ_i\}$ for integer  $r_i, s_i$. 
This will in particular apply to the binary interactions $J_{ij}\in\{0, \pm J\}$ that we are considering in this manuscript.  
 In this case the recursive procedure  Eq.\,\eqref{eq: recursive} requires to evaluate the $m_i(\ell,\tilde{h})$ only on a discrete grid of $(\ell,\tilde h)$ values
 as illustrated in Fig.\,\ref{fig:dynamic programming}. 
 According to Eq.\,\eqref{eq: recursive},  $f_i(\ell,\tilde{h})$ only depends on $f_i(\ell+1,\tilde{h}+J_{i\ell}) $ and $f_i(\ell+1,\tilde{h}) $. The values of the auxiliary field $\tilde h$ at level $\ell+1$ needed to evaluate $f_i(\ell,\tilde{h})$ are thus either the same as at level $\ell$, or differ by $\pm J$ depending on the sign of the coupling $J_{i\ell}$. The arrows in Fig.\, \ref{fig:dynamic programming} are  horizontal in the first case, and diagonally up or down in the second.  The total number of values $f(\ell,\tilde{h})$ that need to be evaluated for a node with $k_i$ predecessors corresponds to the number of markers in Fig.\,\ref{fig:dynamic programming}, and is given by
\begin{equation}
    \sum_{\ell=1}^{k_i+1} \ell = (k_i+1)(k_i+2)/2\ .
\end{equation}
The computational and memory requirement for the evaluation of $f_i$ is thus seen to scale as $k_i^2$, for a node with in-degree $k_i$, rather than 
$2^{k_i}$, as it would in a naive evaluation.
 This entails that the complexity of our algorithm is $\mathcal{O}{(\sum_i k_i^2)}$. 
 If, on the other hand, the  
 couplings $J_{ij}$ were  
 sampled from a continuous distribution, then the function calls would generally be unique, and no speedup would be
 obtained. In this latter case, the complexity remains $\mathcal{O}(\sum_i 2^{k_i})$, making this problem strongly NP-complete \cite{wojtczak2018strong}.
 
 In the following, 
we take full advantage of the computational speedup 
provided by the algorithm discussed above to  
investigate heterogeneities in
Boolean networks with dynamics Eq.\,\eqref{eq: model}, fat-tailed 
degree distribution and thermal noise.

\section{Applications}
\label{sec:applications}
Literature 
on dynamical properties of Boolean or spin systems has traditionally focused on averages of extensive quantities
in the thermodynamic limit \cite{kobayashi2021dynamics,crisanti1988dynamics}. However, the cavity method has been recently used to assess the heterogeneous behaviour of individual nodes \cite{KuRog17,lokhov2014inferring}. These arise from variations in node environment, a prominent feature of networks with fat-tailed degree distributions, and they
have an important role in network processes, as already known in the context of resilience of collective phenomena in networked systems \cite{Albert2000,Annibale+10}. 
However, the aforementioned exponential complexity in the in-degrees
has so far prevented a detailed study of node heterogeneities, precisely in those systems where they play a bigger role.

\begin{figure*}[t]
\includegraphics[width =\textwidth]{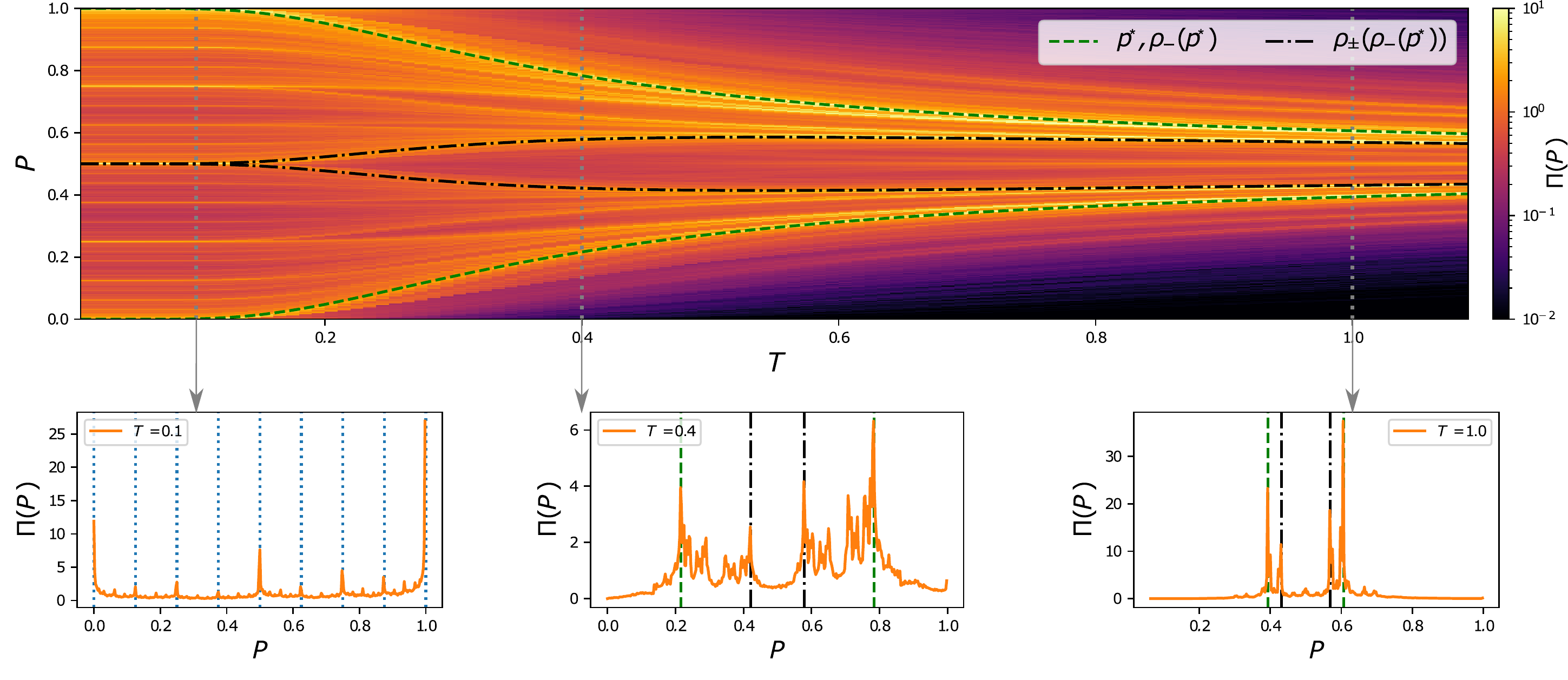}
\caption{(Top) Heat-map of the pdf $\Pi(P)$ of heterogeneous gene activation probabilities at different values of noise parameter $T$. 
The dashed lines show the values $p^\star$ and $\rho_-(p^\star)$, while the dot dashed lines show the values $\rho_\pm(\rho_-(p^\star))$, 
as a function of the noise level $T$. The vertical dotted lines indicate the values of $T$ at which the histograms of $\Pi(P)$, in the lower panels, are computed. 
(Bottom) Histograms of $\Pi(p)$ at different noise levels, i.e. $T=0.1$ (left), $T=0.4$ (middle), $T=1$ (right), 
corresponding to the regimes $T\ll J$, $T\sim J$ and $T\gg J$, respectively. 
The vertical dotted lines on the left panel are located at $(0,1/8,1/4,3/8,\dots 1)$.
 The interactions bias and strength are $\eta=0.621$ and $J \approx 0.72$, respectively. The network size is $N=200,000$. Largest degree of the network is $k = 800$.}
\label{fig: power_law}
\end{figure*}

Here we show that such an analysis is feasible using our method. As an example, we consider 
 a synthetic network of size $N=200,000$ and
 degree distribution $p(k)\sim k^{-\gamma}$, for $k\gg 1$ with $\gamma=2.81$ \footnote{The distribution used is 
$p(k)=[k^{-(\gamma-1)}-(k+1)^{-(\gamma-1)}]$}, chosen to match that of
a gene regulatory network (GRN) prototype \cite{han2018trrust}.  
Our analysis uncovers a highly non-trivial distribution of node activation probabilities.
By computing the stationary solution of Eq.\,\eqref{eq: result cavity}, we obtain the node activation 
probabilities $\lbrace P_i\rbrace_{i=1}^N$ in 
stationary conditions. Their distribution, $\Pi(P)= N^{-1}\sum_i\delta(P_i-P)$, 
is shown in Fig.\,\ref{fig: power_law} for different values of the noise strength $T$. 
For convenience, we fix the threshold at $\vartheta =0$. Thus, the probability $p_0$ that a node spontaneously activates when the local field vanishes, is $p_0 = 1/2$. 
At low noise level, $T\ll J$, 
nodes with frozen neighbours and positive (negative) local field will be active (inactive) with probability very close to one.
This leads to large clusters of nodes locked either in the active or in the inactive state. Correspondingly, the distribution $\Pi(P)$ shows two main peaks at $P=1$ and $P=0$, the former higher than the latter due to the presence of a bias $\eta=0.621$ towards positive interactions (Fig.\ref{fig: power_law}, 
bottom left panel). 
%
The other peaks
correspond to nodes which 
fluctuate between the active and inactive 
state, and thus 
they arise either from nodes with frozen neighbours and zero field or from nodes with fluctuating neighbours. 
In particular, the peak at $P=1/2$ arises from 
nodes that activate spontaneously in a field frozen at zero, while the other peaks arise from nodes with fluctuating neighbours. 
For example, a node $i$ with unit in-degree and neighbour $j$ that is active with probability $p_0$, will be itself active with probability $p_0(1-p_0)$, if the coupling $J_{ij}$ is negative, and 
with probability $p_0+p_0(1-p_0)$ if $J_{ij}$ is positive, as the probability of activation in negative (positive) field is zero (one) respectively. 

When the noise level is increased to $T\sim J$
there is a finite probability that a node activates 
in a negative field. Let us denote 
$\rho_\pm(p)=\frac{1}{2}\left( 1\pm p\tanh\frac{\beta J}{2}\right)$ the activation probability of a node with unit in-degree linked to a neighbour that is active with probability $p$, through the coupling 
$\pm J$, respectively.
Because of the fat tailed degree distribution and absence of correlations 
between in-degrees and out-degrees,
a significant proportion of nodes are found in unidirectional chains (which may and often will be embedded in branching trees). In any such chain, a node $i+1$, successor of a node $i$ that is active with probability $p_i$, will be itself active with 
probability $p_{i+1}=\rho_{\pm}(p_i)$. 
Hence, node activation probabilities are determined by a discrete 
stochastic map with two branches, 
and
two stable fixed points $p^\star=\rho_{+}(p^\star)$ and $p^{\circ}=\rho_{-}(p^{\circ})$. Along the nodes of a chain connected through positive (negative) interactions the activation will converge to $p^{\star}$ ($p^{\circ}$). 
Due to the bias $\eta$, chains of positive interactions are a common occurrence, and yield the rightmost peak of $\Pi(p)$, $P=p^\star$ (see bottom mid and right panels). The leftmost peak arises from nodes linked through a negative coupling to a single node with activation probability $p^\star$, which will activate with probability $\rho_{-}(p^\star)$. 
The other peaks can be interpreted in terms of further
compositions of the functions $\rho_\pm$. 
This in particular also explain the peaks found in the zero-temperature limit at sums of higher order 
powers of $1/2$. 
While the above line of reasoning explains locations of prominent peaks in $\Pi(P)$, other features of the distribution cannot be explained in terms of chain-like structures alone. This includes, in particular, the {\em height\/} of these peaks and {\em any\/} feature of $\Pi(P)$ outside the interval $[\rho_-(1),\rho_+(1)]$, as well as properties of the smooth contributions 
to $\Pi(P)$ observed at finite temperature.
Simulation results are provided in the Supplementary material 
for a network with $N=50,000$ and show excellent agreement 
with predictions from the cavity equations solved via dynamic programming.

\section{Conclusions}
\label{sec:conclusions}
In this manuscript, we have presented a dynamic programming algorithm that dramatically reduces the computational complexity of the dynamic cavity algorithm for Boolean systems with fully asymmetric interactions. It can be used whenever couplings are randomly sampled from (or can be approximated in terms of) a discrete set of equidistant values.
 For systems of this type, the dependence of the computational complexity on the local in-degree of nodes that is required for dynamic updates is reduced from exponential to quadratic. Furthermore, evaluations can be carried out {\it in parallel} for each node.
We illustrate the power of the algorithm by studying a Boolean linear threshold model with binary $\pm J$ interactions. However it can be shown that the present approach can be generalized to non-linear threshold models involving higher order interactions and to bipartite systems, which are of particular interest in the context of gene regulatory networks.
Thanks to the reduction of the computational complexity, we can fully uncover heterogeneities in node activation patterns in 
networks with fat-tailed in-degree distribution,
that were previously inaccessible. 
We discuss salient features of the distribution of node activation probabilities and we show that they can in part be rationalised in terms of discrete stochastic maps resulting from the underlying network structure and dynamics. 
We will report on further aspects of the present model and some of its extensions 
in a future extended paper.

\section{Acknowledgements}
GT is supported by the EPSRC Centre for Doctoral Training in Cross-Disciplinary Approaches to Non-Equilibrium Systems (CANES EP/L015854/1). We thank  prof. Franca Fraternali's group for providing access to their computational facilities.

\appendix
\section{Simulations}
\subsection{Parameter definition}
In the present paper, we consider a  model of a synthetic  network in the configuration model class, i.e. we study ensembles of maximally random directed graphs ${\bf J}$ with constrained in-degree and out-degree sequences. The sequence of in-degree and out-degree are chosen to be the same for simplicity, but the values associated to nodes are shuffled, therefore there is no correlation between the in- and out- degree. The sequence of in-degree of nodes   $k_i^{\rm in}$ is drawn from the distribution $\rho^{\mathrm{in}}(k)
= N^{-1}\sum_i \delta_{k,k_i}$.  We use $\langle k\rangle$ to denote the mean in-degree.  The parameters of our synthetic model are chosen in such a way to match properties of gene regulatory networks, specifically the network described in the dataset  TRRUST \cite{han2018trrust}. 
The in-degree node distribution  is well described by  $P_\gamma(k)= k^{-\gamma}-(k+1)^{-\gamma}$ with power law tail $P_\gamma(k) \sim \gamma k^{-\gamma-1}$ and  $\gamma=2.81$ \cite{han2018trrust}.  
For each non-zero entry of the adjacency matrix $A_{ij}$, the corresponding interaction term $J_{ij}$ can be positive or negative. We sample the sign of the interaction as an independent random variable  with bias  $\eta = 0.621$ reflecting the bias of regulatory interactions (excitatory vs. inhibitory) in TRUST \cite{han2018trrust}. The absolute value of the interaction term is  $J=1/\sqrt{\langle k^{\mathrm{in}}\rangle}$. The size of the network used in the simulation of Fig\,\ref{fig: power_law} is $N = 200,000$.

\subsection{Comparison between theory and simulation }
\begin{figure*}[t]
\begin{center}
\includegraphics[width = 0.48\textwidth]{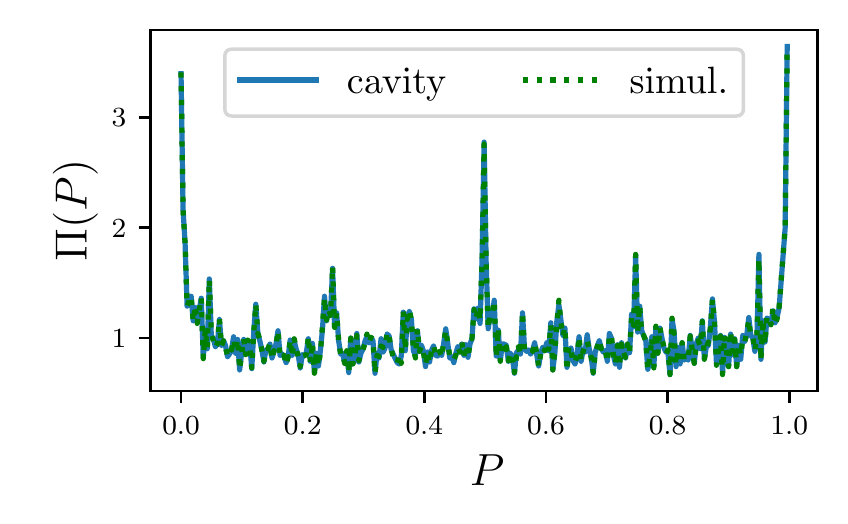}
\includegraphics[width = 0.48\textwidth]{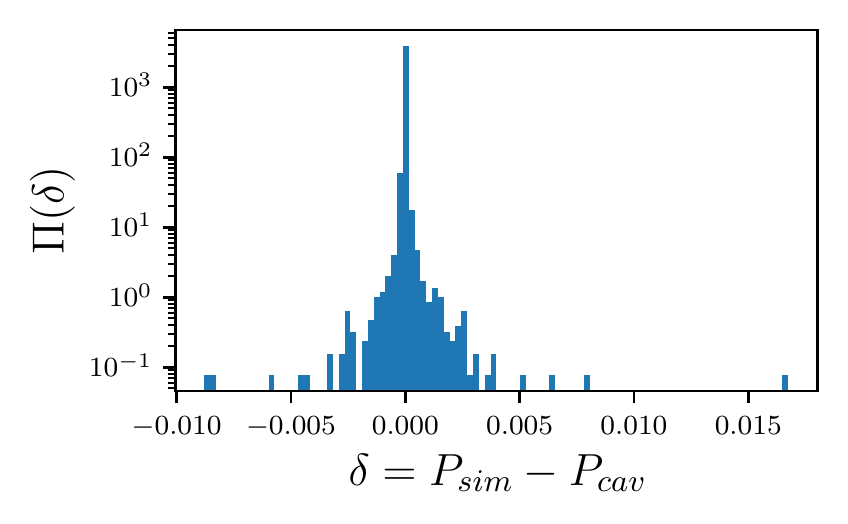}
\end{center}
\caption{Left: Histogram of the node activation probability, as computed from theoretical calculation (blue solid line) and from direct simulation of the microscopic dynamics (dotted green line). Right: Histogram of the difference $\delta_i = P_{i,sim}- P_{i,cav}$, evaluated for each node $i$ of the network.  Parameters: $N =50,000, T=0.1,\gamma=2.81,\eta=0.5$. }
\label{fig:cav_vs_sim}
\end{figure*}

We compute iteratively the node activation probability at time $t+1$ given the knowledge of node activation probabilities at time $t$,
\begin{equation}
P_i(t+1) = \frac{1}{2}\left( 1+ \left\langle \tanh{\left[\frac{\beta}{2}\left(h_i(\bm{n}_{\partial_{i}}) -\vartheta_i\right)  \right]} \right\rangle_{\bm{n}_{\partial_{i}},t}\right)\,.
\label{eq: result cavity_tanh}
\end{equation}
We inspect the stationary solution of the node activation probability $P_i = \lim_{t\rightarrow \infty}P_i(t)$ $\forall i \in\lbrace 1,\dots N\rbrace $.  The stationary activation probabilities    are obtained by running the iterative procedure of Eq.\,\eqref{eq: result cavity_tanh}  until convergence. The convergence is controlled by measuring  the error between consecutive steps $\epsilon_{t+1}=\mathrm{max}_i |P_i(t+1)-P_i(t)| $. Let us call $P_{i,cav}=P_i(t)$ corresponding to the smallest $t$ satisfying the exit condition $\epsilon_t<10^{-4}$.  
We test cavity predictions  against a simulation of the microscopic dynamics Eq.\,\eqref{eq: model}.
 We evaluate the frequency of node activation from dynamical trajectories taking the sample average of node activation\begin{equation}
P_{i,sim}= \frac{1}{t_m}\sum_{t={t_0}}^{t_0+t_{m}}n_i(t)\,,
\label{eq:sim_trj}
\end{equation}
for $t_{m}\gg 1$, and $t_0$ a sufficiently long time to allow the dynamics to relax to stationarity.  In Fig.\,\ref{fig:cav_vs_sim} we show the distribution of the node activation probability as computed by the cavity method and by estimation from simulation. These are found to be in perfect agreement, confirming the validity of the cavity method to investigate the stationary state, see Fig.\,\ref{fig:cav_vs_sim}. Moreover, in order to reach the resolution imposed by cavity $\epsilon_t<10^{-4}$, we need to simulate long trajectories in Eq.\, \eqref{eq:sim_trj} with $t_m\sim10^{8}$, which makes the evaluation from simulation  more computational expensive. For the same network realisation and parameter choice,  the cavity implementation took \SI{50}{\second}, while simulations took more than \SI{11} days on the same machine.
\end{document}